\documentclass[11pt]{article}
\usepackage{latexsym}
\usepackage{amssymb}
\usepackage[dvips]{graphicx}
\usepackage{epsfig}
\usepackage{rotating}

\usepackage{amsfonts}
\usepackage{amsmath}
\begin{document}

\begin{titlepage}
\begin{flushright}
NITheP-09-09\\
ICMPA-MPA/2009/20\\
\end{flushright}

\begin{center}

{\Large\bf Ladder operators and coherent states for continuous spectra
}

Joseph Ben Geloun$^{a,c,d,*}$ and John R. Klauder$^{b,\dag}$

$^{a}${\em National Institute for Theoretical Physics (NITheP)}\\
{\em Private Bag X1, Matieland 7602, South Africa}\\
$^{b}${\em Department of Physics and
Department of Mathematics}\\
{\em University of Florida, Gainesville, FL 32611-8440}\\
$^{c}${\em International Chair of Mathematical Physics
and Applications}\\
{\em ICMPA--UNESCO Chair, 072 B.P. 50  Cotonou, Republic of Benin}\\
$^{d}${\em D\'epartement de Math\'ematiques et Informatique}\\
{\em  Facult\'e des Sciences et Techniques, Universit\'e Cheikh Anta Diop, Senegal}

E-mails:  $^{*}$bengeloun@sun.ac.za,\quad $^{\dag}$klauder@phys.ufl.edu

\begin{abstract}
The notion of ladder operators is introduced for systems with continuous spectra.
We identify two different kinds of annihilation operators allowing
the definition of coherent states as modified ``eigenvectors'' of these operators.
Axioms of Gazeau-Klauder are maintained throughout the construction.
\end{abstract}

\today
\end{center}
Pacs number 03.65.-w

\end{titlepage}

\section{Introduction}
\label{sect:intro}

Coherent states are well known objects with a wide
spectrum of application in mathematics as well as theoretical physics
\cite{kl}-\cite{kl7}. They are generally defined as set of vectors
belonging to a formal Hilbert space, constrained to obey
a set of axioms, that, for the present analysis, we refer to as Gazeau-Klauder
(GK) axioms \cite{kl6}.
Let us recall, as a matter of clarity, this set of suitable requirements.
Given a Hilbert space  ${\mathcal H}$
and a Hamiltonian operator $H$, a system of coherent states of  ${\mathcal H}$,
say $\{|J,\gamma\rangle \}$, is labeled by two real quantities $(J,\gamma)$,  $J\geq 0$,
$\gamma\in \mathbb{R}$, and satisfies the following conditions:
continuity in labels $(J,\gamma)$; resolution of the identity
$\mathbb{I}=\int {\rm d}\mu(J,\gamma)|J,\gamma\rangle \langle J,\gamma|$;
 temporal stability: $e^{- it H}|J,\gamma\rangle = |J,\gamma+\omega t\rangle$
for some constant $\omega$; and the action identity, i.e., $\langle J,\gamma| H| J,\gamma\rangle =\omega J$.
In reference \cite{kl6}, it has been shown that coherent states fulfilling
the GK axioms, can be defined for systems with either discrete, continuous
or both discrete and continuous spectra.

A method for constructing coherent states for systems with discrete spectrum,
is provided by the Barut-Girardello eigenvalue problem for an annihilation operator
with a lowering action on the discrete basis.
As a specific instance, resolving the problem
$a|z\rangle=z|z\rangle$ for the ordinary annihilation operator $a$
satisfying with its adjoint $a^\dag$, the commutation relation
$[a,a^\dag]=\mathbb{I}$, and $z$ a complex variable, leads to the
usual coherent states of the discrete Fock Hilbert space $\{|n\rangle\}$
for the harmonic oscillator. However, the notion
of an annihilation operator onto a continuous basis, is, to the
best of our knowledge, not defined. Other issues arise immediately by consistency.
Even if such an annihilation operator exists, will the resolution of eigenvalue problem
for this operator lead to a system of coherent states? Finally, if
it does, is this set of coherent states the same as the one introduced
by GK in \cite{kl6}?

In this paper, we address these above mentioned issues and
find the following answers. It appears possible to identify
at least two simple types of ladder operators for a system with continuous spectrum,
invoking translation or dilatation\footnote{These two generic situations, from which
the present study is realized, also suggest the definition of a mixed type of
annihilation operator which is not treated here.}
transformations of the continuous parameter labeling a continuous spectrum.
We solve separately the modified ``eigenvalue problem'' generated by each kind of operator,
and show that the resulting states satisfy the GK axioms, and so
can be legitimately called coherent states. Moreover, these coherent states
reduce to those of GK for a particular set of parameter.
Discussions on adjoint operators associated with these annihilation
operators is provided, and we show that these operators obey a deformed Heisenberg algebra.

The paper is organized as follows. In Section 2, we discuss
the first type of an annihilation operator invoking a translation in the continuous
parameter of the Hilbert space basis and the associated set coherent states solutions
of an eigenvalue problem. In Section 3, a similar study is
performed for an annihilation operator involving a dilatation
of the continuous parameter. Section 4 is devoted to concluding remarks
and a short appendix lists some formulas.

\section{Annihilator of the first kind and associated
coherent states}
\label{sect1}

Let us consider a Hamiltonian operator $H>0$ with a nondegenerate continuous
spectrum, and let $|E\rangle$ denote the eigenbasis for this operator,
namely
\begin{eqnarray}
H   |E\rangle = \omega E \,|E\rangle,\;\; 0 < E;  \;\;\;\;\;\;
\langle E|E'\rangle = \delta( E-E').
\end{eqnarray}
Units such that $\hbar=1$ are used.  We will restrict ourself to a
system with an infinite spectrum such that $E\in (0, +\infty)$.

Next, given a real parameter $\varepsilon >0$, we introduce the following
operator
\begin{eqnarray}
a_{\varepsilon}  = \int_0^\infty \;C(E,\varepsilon)  \,|E-\varepsilon\rangle\,\langle E| \;dE,
\label{tran}
\end{eqnarray}
where $C(E,\varepsilon)$ is a free function to be specified
satisfying the condition $C(E,\varepsilon)=0$, for all $0<E< \varepsilon$.
A quick inspection, shows that, for any state $|E\rangle$ with $E-\varepsilon\geq 0$,
$a_{\varepsilon} \,|E\rangle=C(E,\varepsilon)|E-\varepsilon\rangle$.
We will come back soon to the possibility of $\varepsilon=0$ and, later,
the adjoint operator corresponding to $a_{\varepsilon}$ will be discussed.

Let us introduce the states $|s,\gamma\rangle_{\varepsilon}$, $s\in[0,+\infty)$ and
$\gamma\in (-\infty,+\infty)$, by the eigenvalue problem
\begin{eqnarray}
a_{\varepsilon} |s,\gamma\rangle_{\varepsilon} =
(se^{-i\gamma})^\varepsilon\;|s,\gamma\rangle_{\varepsilon} .
\label{egvpb}
\end{eqnarray}
From the present point out view, in analogy with the discrete case,
the usual Barut-Girardello problem (but for a continuous spectrum)
can be recovered for a parameter $\varepsilon=1$ and $z =se^{-i\gamma}$.
The limit $\varepsilon\to 0$ implies that the annihilation operator
(\ref{tran}) is diagonal in the energy; therefore
the eigenvalue $1$ for any state (\ref{egvpb}) will constrain $a_{0}$
to be the identity.

The states  $|s,\gamma\rangle_{\varepsilon}$ can be expanded as
\begin{eqnarray}
|s,\gamma\rangle_{\varepsilon} = \int_0^\infty  \;K_\varepsilon(E;s,\gamma)  \,|E\rangle \;dE.
\end{eqnarray}
The left hand side of (\ref{egvpb})
can be translated (after a change of variable) to become
\begin{eqnarray}
 a_{\varepsilon} |s,\gamma\rangle_{\varepsilon}= \int_0^\infty\;
 C(E+\varepsilon,\varepsilon) K_\varepsilon(E+\varepsilon;s,\gamma)   \,|E\rangle \;dE,
\end{eqnarray}
and, when equated with the right hand side, leads to the functional identity
\begin{eqnarray}
  C(E+\varepsilon,\varepsilon)K_\varepsilon(E+\varepsilon;s,\gamma)=
(se^{-i\gamma})^\varepsilon K_\varepsilon(E;s,\gamma).
\end{eqnarray}
We infer the following relation (after $n$ iterations)
\begin{eqnarray}
\frac{ K_\varepsilon(E+n\varepsilon;s,\gamma)}{K_\varepsilon(E+(n-1)\varepsilon;s,\gamma)}
\ldots
\frac{ K_\varepsilon(E+2\varepsilon;s,\gamma)}{ K_\varepsilon(E+\varepsilon;s,\gamma)}
 \frac{ K_\varepsilon(E+\varepsilon;s,\gamma)}{K_\varepsilon(E;s,\gamma)}=
\frac{\prod_{k=1}^n(se^{-i\gamma})^\varepsilon}{\prod_{k=1}^n C(E+k\varepsilon,\varepsilon)}.
\end{eqnarray}
For the sake of simplicity and without loss of generality,
let us set $E=0$ in the above equation which then leads to
\begin{eqnarray}
K_\varepsilon(n\varepsilon;s,\gamma) =
\frac{(se^{-i\gamma})^{n\varepsilon}}{\prod_{k=1}^n C(k\varepsilon,\varepsilon)}\,
K_\varepsilon(0;s,\gamma) .
\label{sol}
\end{eqnarray}
Before going further, an analogue relation for (\ref{sol})
for the discrete spectrum is given by $\varepsilon=1$ and therefore
 $\prod_{k=1}^n C(k,1)$ stands for the generalized factorial that arises for
 the well known nonlinear coherent states.

Fixing a small $\varepsilon =  \Delta E$,
the convergence of the product $\prod_{k=1}^n C(k\varepsilon,\varepsilon)$
as $n\to \infty$ is ensured if and only
if the function $C(k\varepsilon,\varepsilon)$ possesses the behavior
\begin{eqnarray}
C(k\varepsilon,\varepsilon) \simeq 1 + \bar\alpha k \Delta E  + O((\Delta E)^2),\quad
\end{eqnarray}
for a parameter $\bar\alpha$ depending on $\varepsilon =  \Delta E$.
In this way, the infinite product becomes
\begin{eqnarray}
\lim_{n\to \infty}\prod_{k=1}^n C(k\varepsilon,\varepsilon)
\simeq \lim_{n\to \infty}e^{\bar\alpha\sum_{k=1}^n k\Delta E} =
\lim_{n\to \infty} e^{\bar\alpha \frac{n(n+1)}{2} \Delta E}
\simeq \lim_{n\to \infty}  e^{\alpha \frac{n^2}{2} (\Delta E)^2},
\end{eqnarray}
where we introduced $\bar\alpha = \alpha \Delta E$, with $\alpha$ still being a free parameter.
Then sending both $\Delta E \to 0$ and $n\to \infty$, one arrives at
\begin{eqnarray}
\lim_{n\to \infty}\prod_{k=1}^n C(k\varepsilon,\varepsilon)
= e^{\frac{1}{2}\alpha E^2} .
\end{eqnarray}
Hence, we are led to
\begin{eqnarray}
K_\varepsilon(E;s,\gamma)  &=&
\frac{(se^{-i\gamma})^{E}}{ e^{\frac{1}{2}\alpha E^2}}\,
 K_\varepsilon(0;s,\gamma),\\
 C(E,\varepsilon) &=& e^{\alpha(E\varepsilon- \frac{1}{2}\varepsilon^2)},
 \label{can}
\end{eqnarray}
and where the parameter $\alpha$ and $K_\varepsilon(0;s,\gamma)$ parametrize
the remaining freedom.  As a consequence, the eigenstate solutions
of the eigenvalue problem (\ref{egvpb}) have the general form
\begin{eqnarray}
|s,\gamma\rangle_{\varepsilon}
=   N_\varepsilon(s)
\int_0^\infty  \; \frac{s^{E}}{ e^{\frac{1}{2}\alpha E^2}}\,
e^{-i\gamma E}\,|E\rangle \;dE,
\label{css}
\end{eqnarray}
where $N_\varepsilon(s)= K_\varepsilon(0;s,\gamma)> 0$ will play henceforth the role
of the normalization factor.  The states (\ref{css}) coincide with
those determined by GK \cite{kl6} for a given function $f(E)=e^{\frac{1}{2}\alpha E^2}$.
This shows that the eigenvalue problem allows us to uniquely define a set of
coherent states under these circumstances (up to the parameter $\alpha$).

The normalization to unity of the states (\ref{css}) can be achieved by requiring
$_{\varepsilon}\langle s,\gamma |s,\gamma\rangle_{\varepsilon}=1$ from which one
infers, fixing henceforth $\alpha > 0$,
\begin{eqnarray}
(N_\varepsilon(s))^2=
\left[ \int_0^\infty e^{2E {\rm ln}s - \alpha E^2}\,dE\right]^{-1}=
2\sqrt{\frac{\alpha}{\pi}}
e^{-\frac{({\rm ln}s)^2}{\alpha}}\left[
1- \text{erf}\left(\frac{|{\rm ln}s|}{\sqrt{\alpha}}\right)
\right]^{-1},
\label{norma}
\end{eqnarray}
with ${\rm erf}(\cdot)$ being the Gaussian error function (see Appendix). Notice that this
expression fixes the factor
$K_\varepsilon(0;s,\gamma)= N_\varepsilon(s)$ which does not depend on $\gamma$.

Let us check the main GK axioms in a streamlined fashion:
 \begin{enumerate}
\item[(a)] The continuity in labeling $(s,\gamma)$ is obvious.
\item[(b)] The time evolution:
$ e^{-it H} |s,\gamma\rangle_{\varepsilon}= |s,\gamma+\omega t\rangle_{\varepsilon}$.
\item[(c)] The resolution of the identity:
\begin{eqnarray}
 \int_{-\infty}^{+\infty} \frac{d\gamma}{2\pi}\int_0^{+\infty} ds\, \sigma(s)
|s,\gamma\rangle_{\varepsilon\,\varepsilon}\langle s,\gamma |
 =  \int_0^{\infty} ds\, \sigma(s) (N_\varepsilon(s))^2 \int_0^{\infty}
s^{2E} e^{-\alpha E^2} |E\rangle\langle E|dE
\end{eqnarray}
leads to the Stieljes moment problem
\begin{eqnarray}
  \int_0^{+\infty}  ds \; h(s) s^{2E} =    e^{\alpha E^2},\qquad
  h(s) := \sigma(s) (N_\varepsilon(s))^2.
\end{eqnarray}
Introducing a new variable $u={\rm ln}s$, this problem can be rewritten
as
\begin{eqnarray}
  \int_{-\infty}^{+\infty}  du \; \tilde{h}(u) e^{2Eu} =    e^{\alpha E^2},
\end{eqnarray}
with solution $ \tilde{h}(u)=e^{-\frac{1}{\alpha}u^2}/\sqrt{\alpha\pi}$, so
that the final measure integrating to unity for the system of coherent states
(\ref{css}) is given by
\begin{eqnarray}
\sigma(s) = \frac{1}{s\sqrt{\alpha\pi}} e^{-\frac{1}{\alpha}({\rm ln}s)^2}
(N_\varepsilon(s))^{-2}.
\label{mesur}
\end{eqnarray}
\item[(d)] The action identity can be deduced from the
Hamiltonian mean value:
\begin{eqnarray}
\tilde{H}(s)=\langle s,\gamma| H| s,\gamma\rangle =\omega
(N_\varepsilon(s))^{2}\int_0^\infty \frac{s^{2E}}{e^{\alpha E^2}}
E \; dE =: \omega J(s),
\label{mean}
\end{eqnarray}
where the new action variable $J(s)$ is assumed to be invertible
versus $s$. As argued in \cite{kl6}, if the function
$\tilde{H}(s)/\omega= J(s)$ is invertible (such a condition can be reached
by a strictly increasing or decreasing function $\tilde{H}(s)$, $\tilde{H}'(s)>0$
or $\tilde{H}'(s)<0$) such that $s(J)$ can be determined,
 then the coherent states
$| J,\gamma\rangle :=| s(J),\gamma\rangle$, fulfill all axioms
of GK, and in particular are subjected to the action
identity:
$\langle J,\gamma| H| J,\gamma\rangle
= \langle s(J),\gamma| H| s(J),\gamma\rangle=\omega J$.
The sign of  $\tilde{H}'(s)$ can be tuned by the remaining freedom parameterized
by $\alpha$. It can be shown that for some values of $\alpha>0$, $\tilde{H}'(s)>0$
($\tilde{H}'(s)$ is given in the Appendix).

\end{enumerate}
We have finally succeed to show that the eigenvalue problem (\ref{egvpb})
admits (\ref{css}) as eigenvectors, which are, moreover, coherent states of the GK type.

Let us now determine the adjoint operator
associated with $a_\varepsilon$ and derive an interesting property
satisfied by these operators. A simple Hermitian conjugation allows us
to write
\begin{eqnarray}
 a_\varepsilon^\dag =
\int_0^\infty \;C^*(E,\varepsilon)  \,|E \rangle\,\langle E-\varepsilon|  \;dE
=  \int_0^\infty \;C^*(E+\varepsilon,\varepsilon)  \,
|E +\varepsilon\rangle\,\langle E| \;dE,
\end{eqnarray}
where $C^*(E,\varepsilon)= C(E,\varepsilon)$ is again given by (\ref{can}).
The operators $a_\varepsilon$ and $ a_\varepsilon^\dag$ have the following
algebra
\begin{eqnarray}
{\mathcal I}(\alpha, \varepsilon)&=&
[a_\varepsilon,  a_\varepsilon^\dag ] = \int_{0}^\infty
\left(|C(E+\varepsilon,\varepsilon)|^2 -|C(E,\varepsilon)|^2 \right) \,|E \rangle \langle E| \,
dE  \cr
&=&  \int_{0}^\infty
e^{2\alpha E\varepsilon-\alpha\varepsilon^2}(e^{2\alpha\varepsilon^2}-1)
 \,|E \rangle \langle E| \, dE
\end{eqnarray}
which is a diagonal operator in the energy eigenbasis (therefore commutes
with the energy operator) and consists in a deformed version of the Heisenberg algebra.
Indeed, one recovers the quantum Hilbert space unity $\mathbb{I}$ at the limit
\begin{eqnarray}
\lim_{\alpha\to 0}\frac{{\mathcal I}(\alpha, \varepsilon)}{2\alpha\varepsilon^2}
=\mathbb{I}.
\end{eqnarray}

\section{Annihilator of the second kind and associated
coherent states}
\label{sect2}

In this section, we discuss a second kind of annihilation operator
introduced by a scaling of the parameter $E$ of the continuous Hilbert basis.
The states resolving a problem built out of the annihilation operator
are also shown to be of the GK type.
To emphasize the partial similarity in construction, we use the same notation as used in the previous section
although the quantities may differ.

To proceed with the analysis, we define the operator
\begin{eqnarray}
a^{\lambda}  = \int_0^\infty \;C(E,\lambda)  \,|\lambda E\rangle\,\langle E| \;dE,
\label{dilan}
\end{eqnarray}
where $0< \lambda < 1$ is a real positive
parameter\footnote{In fact, nothing prevents one to choose $\lambda\geq 1$;
the choice $\lambda\in(0,1]$ may be considered analogous to $a_\varepsilon$, in view of
its annihilation (lowering) action, in order to obtain a state $|\lambda E\rangle$
with a label $\lambda E < E$.},
$C(E,\lambda)$ parametrizes the freedom in the definition of $a^{\lambda}$
still to be specified such that at the limit $C(0,\lambda)=0$.
For any state $|E\rangle$, we have
$a^{\lambda} \,|E\rangle=C(E,\lambda)|\lambda E\rangle$; additional
discussion of the adjoint $(a^{\lambda})^\dag$ will follow.

Built differently in comparison to the previous case,
we introduce the states $|s,\gamma\rangle_{\lambda}$, $s\in[0,+\infty)$ and
$\gamma\in (-\infty,+\infty)$, through a new $\lambda$-class of problems:
\begin{eqnarray}
a^{\lambda} |s,\gamma\rangle_{\lambda}=
\frac{1}{\lambda}s^{{\rm ln}\frac{1}{\lambda}}\;|s,\frac{\gamma}{\lambda}\rangle_{\lambda} .
\label{egvpb2}
\end{eqnarray}
It is worth noting that this problem is not an eigenvalue problem.
In addition, the usual eigenvalue problem cannot be obtained
for any value of $\lambda$. However, two specific cases have to be discussed:
the limit $\lambda=1$ and the situation $\lambda=e^{-1}$.
These generate problems of the form
$a^{1} |s,\gamma\rangle_{1}=|s,\gamma\rangle_{1}$
and
$a^{e^{-1}} |s,\gamma\rangle_{e^-1}=es|s,e\gamma\rangle_{e^-1}$, respectively.
Sending $\lambda\to 1$, the annihilator (\ref{dilan}) is a diagonal operator
with an eigenvalue $1$; this clearly constrains $a^{1}$ to be the identity.
For $\lambda=e^{-1}$, we are led  very close to --
but still continuous and thus different from -- the ordinary Barut-Girardello problem.

The states  $|s,\gamma\rangle_{\lambda}$ can be expanded in the continuous
basis as
\begin{eqnarray}
|s,\gamma\rangle_{\lambda} = \int_0^\infty  \;K_\lambda(E;s,\gamma)  \,|E\rangle \;dE,
\end{eqnarray}
with $K_\lambda(E;s,\gamma)$ being complex coefficients.
The first member of (\ref{egvpb2}) can be put in the form
\begin{eqnarray}
a^{\lambda} |s,\gamma\rangle_{\lambda}= \int_0^\infty\;
\frac{1}{\lambda} C(\frac{E}{\lambda},\lambda)\; K_\lambda(\frac{E}{\lambda}; s,\gamma)
 \,|E\rangle \;dE,
\end{eqnarray}
and, when equated with the second member, gives
\begin{eqnarray}
  C(\frac{E}{\lambda},\lambda)K_\lambda(\frac{E}{\lambda};s,\gamma)=
 s^{{\rm ln}\frac{1}{\lambda}}\, K_\lambda(E;s,\frac{\gamma}{\lambda}).
\end{eqnarray}
We will assume a separation of the variables $s$ and $\gamma$
in term of the ansatz $K_\lambda(E;s,\gamma)= K^0_\lambda(E;s)e^{-i\gamma E}$,
such that the phase function will reproduce both the correct time evolution
of these states and the relation
$e^{-i\gamma \cdot \frac{E}{\lambda}}=e^{-i\frac{\gamma}{\lambda} \cdot E}$.
Factoring out this phase contribution, one gets
\begin{eqnarray}
  C(\frac{E}{\lambda},\lambda)K^0_\lambda(\frac{E}{\lambda};s)=
 s^{{\rm ln}\frac{1}{\lambda}}\, K^0_\lambda(E;s)
 \label{iter}
\end{eqnarray}
which can be solved along the lines of the previous analysis.
First, let us introduce $\tilde{K}_\lambda({\rm ln}E;s)=K^0_\lambda(E;s)$
and $\tilde{C}({\rm ln}E,\lambda)=C(E,\lambda)$.  By iteration from (\ref{iter}),
we find that
\begin{eqnarray}
\frac{ \tilde{K}_\lambda({\rm ln}E - n{\rm ln}\lambda;s)}{\tilde{K}_\lambda
({\rm ln}E-(n-1){\rm ln}\lambda;s)}
\ldots
\frac{ \tilde{K}_\lambda({\rm ln}E - 2{\rm ln}\lambda;s)}{
\tilde{K}_\lambda({\rm ln}E - {\rm ln}\lambda;s)}
 \frac{ \tilde{K}_\lambda({\rm ln}E - {\rm ln}\lambda;s)}{
 \tilde{K}_\lambda({\rm ln}E;s)}=
\frac{s^{-n {\rm ln}\lambda}}{
\prod_{k=1}^n \tilde{C}({\rm ln}E - k{\rm ln}\lambda,\lambda)}
\end{eqnarray}
and setting $E=1$, it follows that
\begin{eqnarray}
\tilde{K}_\lambda( - n{\rm ln}\lambda;s)
=
\frac{s^{-n {\rm ln}\lambda}}{
\prod_{k=1}^n \tilde{C}(- k{\rm ln}\lambda,\lambda)}  \tilde{K}_\lambda(0;s) .
\end{eqnarray}
The same routine for the convergence of the infinite product as
$n\to \infty$ and with small ${\rm ln}\lambda =  \Delta \tilde{\tilde{E}}=\Delta({\rm ln}E)$,
requires that
\begin{eqnarray}
 \tilde{C}(- k{\rm ln}\lambda,\lambda) \simeq 1 + \bar\beta k \Delta \tilde{\tilde{E}}
  + O((\Delta \tilde{\tilde{E}})^2),
\end{eqnarray}
with $\bar\beta$ depending on ${\rm ln}\lambda =  \Delta \tilde{\tilde{E}}$.
The infinite product becomes
\begin{eqnarray}
\lim_{n\to \infty}\prod_{k=1}^n  \tilde{C}(- k{\rm ln}\lambda,\lambda)
\simeq \lim_{n\to \infty}e^{\bar\beta\sum_{k=1}^n k \Delta \tilde{\tilde{E}}} =
\lim_{n\to \infty} e^{\bar\beta \frac{n(n+1)}{2}  \Delta \tilde{\tilde{E}}}
\simeq \lim_{n\to \infty}  e^{\beta \frac{n^2}{2} (\Delta \tilde{\tilde{E}})^2}.
\end{eqnarray}
Here $\bar\beta = \beta  \Delta \tilde{\tilde{E}}$ and
$\beta$ is again a free parameter.
In the limit $\Delta \tilde{\tilde{E}} \to 0$ and $n\to \infty$,
we obtain
\begin{eqnarray}
\lim_{n\to \infty}\prod_{k=1}^n  \tilde{C}(- k{\rm ln}\lambda,\lambda)
\simeq e^{\frac{1}{2}\beta {\tilde{\tilde{E}}}^2}= e^{\frac{1}{2}\beta({\rm ln}E)^2} .
\end{eqnarray}
We are then able to identify the functions
\begin{eqnarray}
K^0_\lambda(E;s)  &=&
\frac{s^{{\rm ln}E}}{ e^{\frac{1}{2}\beta ({\rm ln}E)^2}}\,
 K^0_\lambda(1;s)\\
 C(E,\lambda) &=& e^{\beta(({\rm ln}E)({\rm ln}\lambda) - \frac{1}{2}({\rm ln}\lambda)^2)},
 \label{can2}
\end{eqnarray}
with $\beta$ and $K^0_\lambda(1;s)$ free quantities.
Solutions of the problem (\ref{egvpb2}) have the general form
\begin{eqnarray}
|s,\gamma\rangle_\lambda
=   N_\lambda(s)
\int_0^\infty  \; \frac{s^{{\rm ln}E}}{ e^{\frac{1}{2}\beta ({\rm ln}E)^2}}\,
e^{-i\gamma E}\,|E\rangle \;dE,
\label{css3}
\end{eqnarray}
where $N_\lambda(s)= K_\lambda(1;s)> 0$ is the normalization factor.
Comparing these states with those of GK, one ends with the function
$f(E)= e^{\frac{1}{2}\beta ({\rm ln}E)^2}$ uniquely specifying
this set of coherent states.

Insisting on normalizing the states (\ref{css3}), the following
relation holds
\begin{eqnarray}
  (K_\lambda(1;s))^2 =(N_\lambda(s))^2 =
\left[\int_0^\infty\; \frac{s^{2{\rm ln}E}}{ e^{\beta ({\rm ln}E)^2}}\;dE\right]^{-1}
=\sqrt{\frac{\beta}{\pi}} e^{-\frac{(2 {\rm ln}s+1)^2}{4 \beta }} .
\label{norm}
\end{eqnarray}
The GK axioms can also be explicitly verified. Omitting the proof of the
continuity in labeling and correct time evolution, both easily obtained,
let us address the resolution of the identity. We have
\begin{eqnarray}
 \int_{-\infty}^{+\infty} \frac{d\gamma}{2\pi}\int_0^{+\infty} ds\, \rho(s)
|s,\gamma\rangle_{\lambda\,\lambda}\langle s,\gamma |
 =  \int_0^{\infty} ds\, \rho(s) (N_\lambda(s))^2 \int_0^{\infty}
s^{2{\rm ln}E} e^{-\beta ({\rm ln}E)^2} |E\rangle\langle E|dE
\end{eqnarray}
inducing the moment problem
\begin{eqnarray}
  \int_0^{+\infty}  ds \; h(s) s^{2{\rm ln}E} =    e^{\beta ({\rm ln}E)^2},\qquad
  h(s) := \rho(s) (N_\lambda(s))^2,
  \label{stj}
\end{eqnarray}
which can be solved as previously by
using the variable  $u={\rm ln}s$.
The solution as a function of $u$ is
$ \tilde{h}(u)=e^{-\frac{1}{\beta}u^2}/\sqrt{\beta\pi}$.
Therefore, the overall measure leading to a resolution of unity for the system of states
(\ref{css3}) is
\begin{eqnarray}
 \rho(s) = \frac{1}{s\sqrt{\beta\pi}} e^{-\frac{1}{\beta}({\rm ln}s)^2}
(N_\lambda(s))^{-2}= \frac{1}{s\beta}e^{-\frac{1}{4\beta}(4{\rm ln}s+1)}
\end{eqnarray}
differing from (\ref{mesur}) by the norm factor,
and hence yielding a new family of coherent states.

The action identity can be inferred from the expression of the
Hamiltonian mean value
\begin{eqnarray}
\tilde{H}(s)=_{\,\lambda}\!\langle s,\gamma| H| s,\gamma\rangle_\lambda =\omega
(N_\lambda(s))^{2}\int_0^\infty \frac{s^{2{\rm ln}E}}{e^{\beta ({\rm ln}E)^2}}
E \; dE =: \omega J(s).
\label{mean3}
\end{eqnarray}
The new action variable $J(s)$ has to be inverted in terms of $s(J)$.
The integration (\ref{mean3}) can be performed exactly; one finds $J(s)$, which
turns out to be explicitly invertible as
\begin{eqnarray}
J(s)=\frac{\tilde{H}(s)}{\omega }= e^{\frac{1}{\beta}({\rm ln}s +\frac{3}{4})},\qquad
s(J) = e^{\beta {\rm ln}J -\frac{3}{4}}.
\label{jl}
\end{eqnarray}
The correct variable in term of which all GK axioms can be reached
is $J$ and the associated coherent states $|J,\gamma\rangle_\lambda
= |s(J),\gamma\rangle_\lambda$ can be written as
\begin{eqnarray}
|J,\gamma\rangle_\lambda
=   N_\lambda(s(J))
\int_0^\infty  \; \frac{(e^{\beta {\rm ln}J -\frac{3}{4}})^{{\rm ln}E}}{
e^{\frac{1}{2}\beta ({\rm ln}E)^2}}\,
e^{-i\gamma E}\,|E\rangle \;dE.
\label{css4}
\end{eqnarray}

Concerning properties of the adjoint operator associated with $a^\lambda$,
we have
\begin{eqnarray}
 (a^\lambda)^\dag =
\int_0^\infty \;C^*(E,\lambda)  \,|E \rangle\,\langle \lambda E |
=\frac{1}{\lambda}
\int_0^\infty \;C^*(\frac{E}{\lambda},\lambda)|\frac{E}{\lambda} \rangle\,\langle E| \;dE,
\end{eqnarray}
with $C^*(E,\lambda)= C(E,\lambda)$ given by (\ref{can2}).
The following $\lambda$-deformed relation holds
\begin{eqnarray}
{\mathcal I}(\beta, \lambda)&=&
[a^\lambda,  (a^\lambda)^\dag]_\lambda :=
a^\lambda (a^\lambda)^\dag -\frac{1}{\lambda}  (a^\lambda)^\dag a^\lambda\cr
&=& \int_{0}^\infty \, \frac{1}{\lambda}
\left(|C^*(\frac{E}{\lambda},\lambda)|^2 -|C(E,\lambda)|^2 \right)
\,|E \rangle \langle E| \,dE  \cr
&=&  \int_{0}^\infty
e^{2\beta ({\rm ln}E)({\rm ln}\lambda) - \beta  ({\rm ln}\lambda)^2}
\frac{(1-e^{2\beta({\rm ln}\lambda)^2})}{\lambda}
 \,|E \rangle \langle E| \, dE,
\end{eqnarray}
which defines a diagonal operator in the energy eigenbasis.
This operator characterizes again a deformed version of the Heisenberg algebra
since in the limit
\begin{eqnarray}
\lim_{\beta \to 0}
\frac{\lambda{\mathcal I}(\beta, \lambda)}{-2\beta({\rm ln}\lambda)^2}=\mathbb{I},
\end{eqnarray}
the ordinary bosonic algebra can be recovered.

\section{Conclusion}
\label{ccl}

We have studied ladder operators for systems with continuous
and infinite spectra. These operators are defined through translation
or dilatation of the continuous parameter labeling a given spectrum.
We have succeed in solving, for both cases and in the continuous limit,
the problems defining coherent states generalizing Barut-Girardello
eigenvalue problem in the discrete case.
The resulting coherent states are different for each kind
of annihilation operator and prove to fulfill all the requirements
of GK, thus enlarging prime classes of coherent states with an exact
resolution of the identity.
Finally, in this construction, we show that new kinds of deformed
Heisenberg algebras are satisfied
by the annihilation operator and its adjoint.

\section*{Acknowledgments}
J.R.K. thanks the National Institute for Theoretical Physics (NITheP)
and its Director Prof. Frederik G. Scholtz for hospitality and support
during a pleasant stay in Stellenbosch. Both authors thank Prof. Jan
Govaerts for a helpful remark regarding our analysis. This work
was supported under a grant of the  National Research Foundation of South Africa.

\section{Appendix}
This appendix lists useful identities.

\begin{enumerate}
\item[(i)] The Gaussian error functions are defined by
\begin{eqnarray}
{\rm erf}(x) = \frac{2}{\sqrt{\pi}}\int_0^x e^{-t^2} dt,\qquad
{\rm erfc}(x) = 1-{\rm erf}(x)=
\frac{2}{\sqrt{\pi}}\int_x^\infty e^{-t^2} dt.
\end{eqnarray}
The derivation of the expression of the norm (\ref{norma}) can be performed
as follows. Fixing $\alpha >0$, we have
\begin{eqnarray}
I(s) = \int_{0}^\infty e^{-\alpha\left(E-\frac{{\rm ln}s}{\alpha}\right)^2
+\frac{({\rm ln}s)^2}{\alpha}} = e^{\frac{({\rm ln}s)^2}{\alpha}}
\int_{-\frac{{\rm ln}s}{\alpha}}^\infty e^{-\alpha X^2} dX.
\end{eqnarray}
Then, two cases may occur: (a) If  $-{\rm ln}s> 0$, then
\begin{eqnarray}
I_+(s) &=& e^{\frac{({\rm ln}s)^2}{\alpha}}
\int_{-\frac{{\rm ln}s}{\alpha}}^\infty e^{-\alpha X^2} dX=
e^{\frac{({\rm ln}s)^2}{\alpha}}
\left\{ \int_{0}^\infty -
\int_0^{-\frac{{\rm ln}s}{\alpha}} \right\} e^{-\alpha X^2} dX\cr
&=&
\frac{1}{2}\sqrt{\frac{\pi}{\alpha}}\left[1 - {\rm erf}\left(-\frac{{\rm ln}s}{\sqrt{\alpha}}\right)\right]
\end{eqnarray}
or (b)  $-{\rm ln}s < 0$, then
\begin{eqnarray}
I_-(s) &=& e^{\frac{({\rm ln}s)^2}{\alpha}}
\int_{-\frac{{\rm ln}s}{\alpha}}^\infty e^{-\alpha X^2} dX=
e^{\frac{({\rm ln}s)^2}{\alpha}}
\left\{ \int_{0}^\infty +
\int^0_{-\frac{{\rm ln}s}{\alpha}} \right\} e^{-\alpha X^2} dX\cr
&=&
\frac{1}{2}\sqrt{\frac{\pi}{\alpha}}\left[1 - {\rm erf}\left(\frac{{\rm ln}s}{\sqrt{\alpha}}\right)\right].
\end{eqnarray}
Finally, for all $s$, we have
\begin{eqnarray}
I(s) = e^{\frac{({\rm ln}s)^2}{\alpha}}
\int_{-\frac{{\rm ln}s}{\alpha}}^\infty e^{-\alpha X^2} dX=
\frac{1}{2}\sqrt{\frac{\pi}{\alpha}}\left[1 - {\rm erf}\left(\frac{|{\rm ln}s|}{\sqrt{\alpha}}\right)\right],
\end{eqnarray}
which has to be inverted before recovering (\ref{norma}).

\item[(ii)] Given the Hamiltonian mean value function $\tilde{H}(s)$ (\ref{mean})
\begin{eqnarray}
\tilde{H}(s)&=&\langle s,\gamma| H| s,\gamma\rangle =\omega
(N_\varepsilon(s))^{2}\int_0^\infty \frac{s^{2E}}{e^{\alpha E^2}}
E \; dE =: \omega J(s),\cr
&=&\omega\;\frac{\sqrt{\pi }\; \text{erfc}\left(\frac{|{\rm ln } (s)|}{\sqrt{\alpha}}\right) {\rm ln }(s)+\sqrt{\alpha}e^{-\frac{{\rm ln }^2(s)}{\alpha}} }{ \sqrt{\pi }\alpha^{\frac{3}{2}}\text{erfc}\left(\frac{|{\rm ln } (s)|}{\sqrt{\alpha}}\right) },
\label{mean2}
\end{eqnarray}
and using the obvious formula $\partial_x {\rm erf} (x) = (2/\sqrt{\pi}) e^{-x^2}$,
we can obtain the derivative $\partial_s \tilde{H}(s)$ as
\begin{eqnarray}
\tilde{H}'_\alpha(s) = \frac{\omega}{\alpha s}\left[-\frac{2 e^{-\frac{{\rm ln }^2(s)}{\alpha}} {\rm ln } (s)}{\sqrt{\alpha\pi } \,\text{erfc}\left(\frac{|{\rm ln } (s)|}{\sqrt{\alpha}}\right)}
+\frac{2 e^{-\frac{2 {\rm ln }^2(s)}{\alpha}}}{\pi  \left(\text{erfc}\left(\frac{|{\rm ln } (s)|}{\sqrt{\alpha}}\right)\right)^2}+1 \right].
\end{eqnarray}
For $\alpha=1$, this expression reduces to
\begin{eqnarray}
\tilde{H}'_1(s) = \frac{\omega}{s}
\left[-\frac{2 e^{-{\rm ln }^2(s)} {\rm ln } (s)}{\sqrt{\pi } \text{erfc}\left(|{\rm ln } (s)|\right)}
+\frac{2 e^{-2 {\rm ln }^2(s)}}{\pi  \left(\text{erfc}\left(|{\rm ln } (s)|\right)\right)^2}+1 \right]\geq 0.
\end{eqnarray}
\end{enumerate}

\end{document}